\documentclass{sigchi}
\usepackage{url} 
\usepackage{float}
\usepackage{xcolor}

\CopyrightYear{2019}
\setcopyright{rightsretained}

\usepackage{balance}       
\usepackage{graphics}      
\usepackage[T1]{fontenc}   
\usepackage{txfonts}
\usepackage{mathptmx}
\usepackage[pdflang={en-US},pdftex]{hyperref}
\usepackage{color}
\usepackage{booktabs}
\usepackage{textcomp}
\usepackage{etoolbox}

\usepackage{microtype}        
\usepackage{ccicons}          

\usepackage{todonotes}

\def\plaintitle{Threshold Detection for Indoor-Localization- and Proximity-aware Applications using Inaudible Audio}
\def\plaintitle{Augmenting Indoor-Localization and Proximity-Detection with Inaudible Audio to Improve Intelligent Applications}
\def\plaintitle{Using Inaudible Audio to Improve Indoor-Localization- and Proximity-Aware Intelligent Applications}

\def\emptyauthor{}
\def\plainkeywords{indoor localization; proximity; inaudible audio}

\makeatletter
\def\url@leostyle{%
  \@ifundefined{selectfont}{
    \def\UrlFont{\sf}
  }{
    \def\UrlFont{\small\bf\ttfamily}
  }}
\makeatother
\def\pprw{8.5in}
\def\pprh{11in}

\definecolor{linkColor}{RGB}{6,125,233}
\hypersetup{%
  pdftitle={\plaintitle},
  pdfauthor={\emptyauthor},
  pdfkeywords={\plainkeywords},
  pdfdisplaydoctitle=true, 
  bookmarksnumbered,
  pdfstartview={FitH},
  colorlinks,
  citecolor=black,
  filecolor=black,
  linkcolor=black,
  urlcolor=linkColor,
  breaklinks=true,
  hypertexnames=false
}

\begin{document}

\title{\plaintitle}

\numberofauthors{2}
\author{%
  \alignauthor{Scott A. Carter, Daniel Avrahami\\
    \affaddr{FX Palo Alto Laboratory, Inc.}\\
    \affaddr{3174 Porter Drive}\\
    \affaddr{Palo Alto, California 94304 USA}\\
    \email{carter,daniel@fxpal.com}}\\
  \alignauthor{Nami Tokunaga\\
    \affaddr{Communication Technology Laboratory}\\
    \affaddr{Research \& Technology Group}\\
    \affaddr{Fuji Xerox Co., Ltd.}\\
    \affaddr{6-1 Minatomirai, Nishi-ku, Yokohama-shi Kanagawa 220-8668 Japan}\\
    \email{nami.tokunaga@fujixerox.co.jp}}\\
}

\maketitle

\begin{abstract}
While it is often critical for indoor-location- and proximity-aware applications to know whether a user is in a space or not (e.g., a specific room or office), a key challenge is that the difference between standing on one side or another of a doorway or wall is well within the error range of most RF-based approaches. In this work, we address this challenge by augmenting RF-based localization and proximity detection with active ultrasonic sensing, taking advantage of the limited propagation of sound waves. This simple and cost-effective approach can allow, for example, a Bluetooth smart-lock to discern whether a user is inside or outside their home in order to lock or unlock doors automatically. We describe a configurable architecture for our solution and present experiments that validate this approach but also demonstrate that different user behavior and application needs can impact system configuration decisions. Finally, we describe applications that could benefit from our solution and address privacy concerns.
\end{abstract}

\category{H.5.m.}{Information Interfaces and Presentation
  (e.g. HCI)}{Miscellaneous}{}{}


\section{Introduction}
Indoor localization and proximity detection are key ingredients for many intelligent applications, services, and smart connected devices. Significant progress has been made in the accuracy and performance of wireless, radio-frequency-based (RF-based) localization solutions, using, for example, GSM \cite{otsason_accurate_2005}, Wi-Fi \cite{bahl_radar:_2000,chintalapudi_indoor_2010,biehl_loco:_2014,li_indotrack:_2017,sugasaki_robust_2017,soltanaghaei_multipath_2018}, UWB \cite{correal_uwb_2003}, Zigbee \cite{sugano_indoor_2006}, and different combinations \cite{cooper_loco:_2016} (for surveys, see \cite{liu_survey_2007,mainetti_survey_2014}). However, despite many advancements, it is still difficult for most RF-based solutions to determine if a tracked device or user is on one side of a wall or door or the other. Even past work that combines RF and inaudible audio tends to work poorly near threshold boundaries (see Figure 9 in \cite{Sosa-Sesma}). This is because the physical distance between standing on one side or another of a physical threshold is well within the accuracy limits of most localization systems. 

Consider, for example, a smart lock that cannot tell for certain whether a user is inside or outside their home, or a private document-delivery service that cannot ascertain that the user is within the designated location or not in the adjacent office.
We propose a simple yet valuable approach for overcoming this challenge by augmenting RF-based localization and proximity-detection with active audio sensing. We focus on \emph{inaudible audio}: sound waves just above 20kHz, which are beyond typical human hearing range \cite{rosen} but that are still within range of commodity microphones. Unlike RF, physical barriers, such as a closed door or walls, more strongly attenuate sound waves, making them particularly useful for threshold-detection. 

In this paper, we present a solution that relies on broadcasting and listening for ultrasonic phrases to allow an indoor-location or proximity-detection system to determine on which side of a threshold a user's device is located. We describe two potential configurations for our solution, each with different technical and privacy characteristics, and present results from an evaluation of our solution under different conditions. Specifically, this paper makes the following contributions: 1) A simple and cost-effective solution to determining position around a threshold; 2) Results from a series of tests that validate the approach but also reveal that application needs as well as the method of carrying a mobile device can have surprising impacts on optimal system configuration, and 3) a discussion of performance as well as privacy trade-offs for this use of acoustic sensing.

\section{Related work}
A growing body of work has demonstrated the broad appeal of acoustic sensing as a low cost and ubiquitous approach for spatial interaction \cite{gupta_soundwave:_2012,aumi_doplink:_2013, watanabe_ultrasound-based_2013,laput_sweepsense:_2016,nishida_real_2004,zhang_soundtrak:_2017}, context-aware applications \cite{li_condiosense:_2017}, localization \cite{rossi_roomsense:_2013,tarzia_indoor_2011,azizyan_surroundsense:_2009,schweinzer_ultrasonic_2010,lazik_indoor_2012,mao_cat:_2016,qiu_silent_2017}, and device-to-device communication and interaction \cite{lee_chirp_2015,tung_cross-platform_2018,wang_messages_2016,Jin2015}. However, many of these systems require extensive infrastructure, which can be difficult to deploy and maintain \cite{Ward}. Other work mentions threshold detection (for automobiles) only as future work \cite{laput_sweepsense:_2016}.

In the area of indoor localization, passive acoustic sensing can be used to match fingerprints of ambient audio captured in different spaces \cite{azizyan_surroundsense:_2009,rossi_roomsense:_2013,tarzia_indoor_2011}. One potentially limiting assumption with this approach is that spaces must be sufficiently acoustically different. With active acoustic sensing-based localization, audio is emitted and captured by the user's device or speakers in the environment \cite{schweinzer_ultrasonic_2010,lazik_indoor_2012,mao_cat:_2016,qiu_silent_2017}. Mao \textit{et al.}, for example, use an array of ultrasonic speakers within a room to perform accurate 3D spatial localization \cite{mao_cat:_2016}. 

For the task of determining what side of a threshold or wall a user is standing, prior research also explored several non-acoustic approaches. For example, sensing changes in air-pressure when people move between rooms or close doors \cite{patel_detecting_2008}, or by using a radar mounted above the doorway and categorizing users by their body shape \cite{kalyanaraman_forma_2017}. Others use knowledge of walls within the space from map data to improve non-acoustic predicted localization (\textit{c.f.}, \cite{li_reliable_2012,lazik_indoor_2012,biswas_wifi_2010}). Our approach, which we describe next, relies on active acoustic sensing as a cost-effective solution for augmenting RF-based localization or proximity-detection systems.

\enlargethispage{\baselineskip}

\section{Augmented Proximity and Location Detection}

Our system uses ultrasonic audio beacons to augment RF-based smart applications. With standard RF-based systems, mobile devices can communicate identifying information to fixed base stations. Once the mobile application receives an RF UUID that matches a known UUID, it generates an ultrasonic message. This message can be static, globally unique, or can incorporate details of the application context, such as the RF UUID or session information. The mobile then sends this ultrasonic (US) message continuously. When the fixed receiver detects an RF signal but no US signal from the mobile beacon, it can infer that the mobile device is on the other side of a threshold. The fixed beacon can relay this information to external services (e.g., a smart door service). Once the fixed receiver detects both the RF signal and the US signal, it can infer that the mobile device is inside the space and again relay this information to external services (e.g., to lock the  door). 

Note that if there are multiple mobile devices in a space, they will all be sending ultrasonic signals to the fixed receiver. To avoid interference, mobile devices can broadcast messages on different frequency channels. The fixed device can assign open channels to each mobile device as they move within RF-range. 

\begin{figure}[tb]
\centering
\begin{tabular}{c} 
\includegraphics[width=3in]{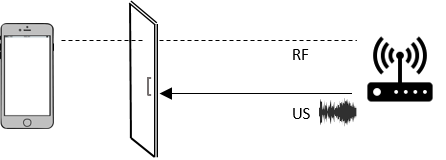} \\
(a) \\
\includegraphics[width=3in]{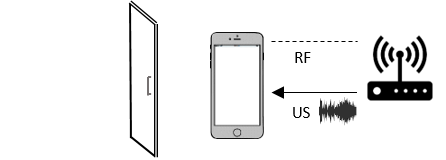} \\
(b) \\
\end{tabular}
\caption{Mobile and fixed devices communicate through RF and ultrasonics (US). (a) The beaconing device (here, the fixed device) then begins emitting ultrasonic messages, which cannot cross thresholds (such as doors or walls). (b) When the fixed and mobile devices are in the same space, the listening device receives ultrasonic messages and can communicate verified presence in the space to external services.}
\label{fig:system}
\end{figure}

Furthermore, any device with a speaker and microphone can serve as either an ultrasonic beacon or a receiver. Therefore, in the scenario describe above, the fixed device could \emph{send} ultrasonic messages and the mobile device could \emph{receive} them (Figure \ref{fig:system}). Since in that case there is only one device transmitting messages (the fixed device) this approach mitigates the need for multi-channel support. On the other hand, it makes the mobile device responsible for determining the user's location.

Finally, it is also possible to completely remove the reliance on RF-based signals. To support that case, a second ultrasonic beacon can be placed on the outside of a threshold. When the mobile device comes within range of the exterior ultrasonic device, the system can communicate to external services that the user is nearby but outside (e.g., to open a smart door). When the user is inside with the door closed, the system will detect ultrasonic signals at the interior fixed receiver but not the exterior fixed receiver, indicating that the user is inside (e.g., to lock the door).

\subsection{Implementation}
A number of tools and frameworks are available for making the development of ultrasonic-based applications simpler (cf., \cite{lee_chirp_2015,tung_cross-platform_2018,wang_messages_2016}).
In our proof-of-concept implementation, we use the Chirp SDK, an open-source framework for sending and receiving audible and inaudible audio with any desktop or mobile device \cite{chirp}. This flexibility allows us to export the system to any device with a microphone and/or a speaker. Also, the system requires no network connection and therefore can work in areas or situations that have no internet connectivity. We have also tested the system in the context of a standard teleconference meeting as well as background music in a cafe setting. In neither case was performance effected. 

Currently, beacons send short messages (two arbitrary hexadecimal characters) and use a delay to detect absence. However, we have successfully tested messages up to 16 hex characters (8 bytes) in length. It is therefore possible to generate keys specific to each application context, which can prevent against potential man-in-the-middle attacks.

\section{Evaluation}

\begin{figure}[tb]
\centering
\begin{tabular}{cccccc} 
\multicolumn{3}{c}{\includegraphics[width=1.5in]{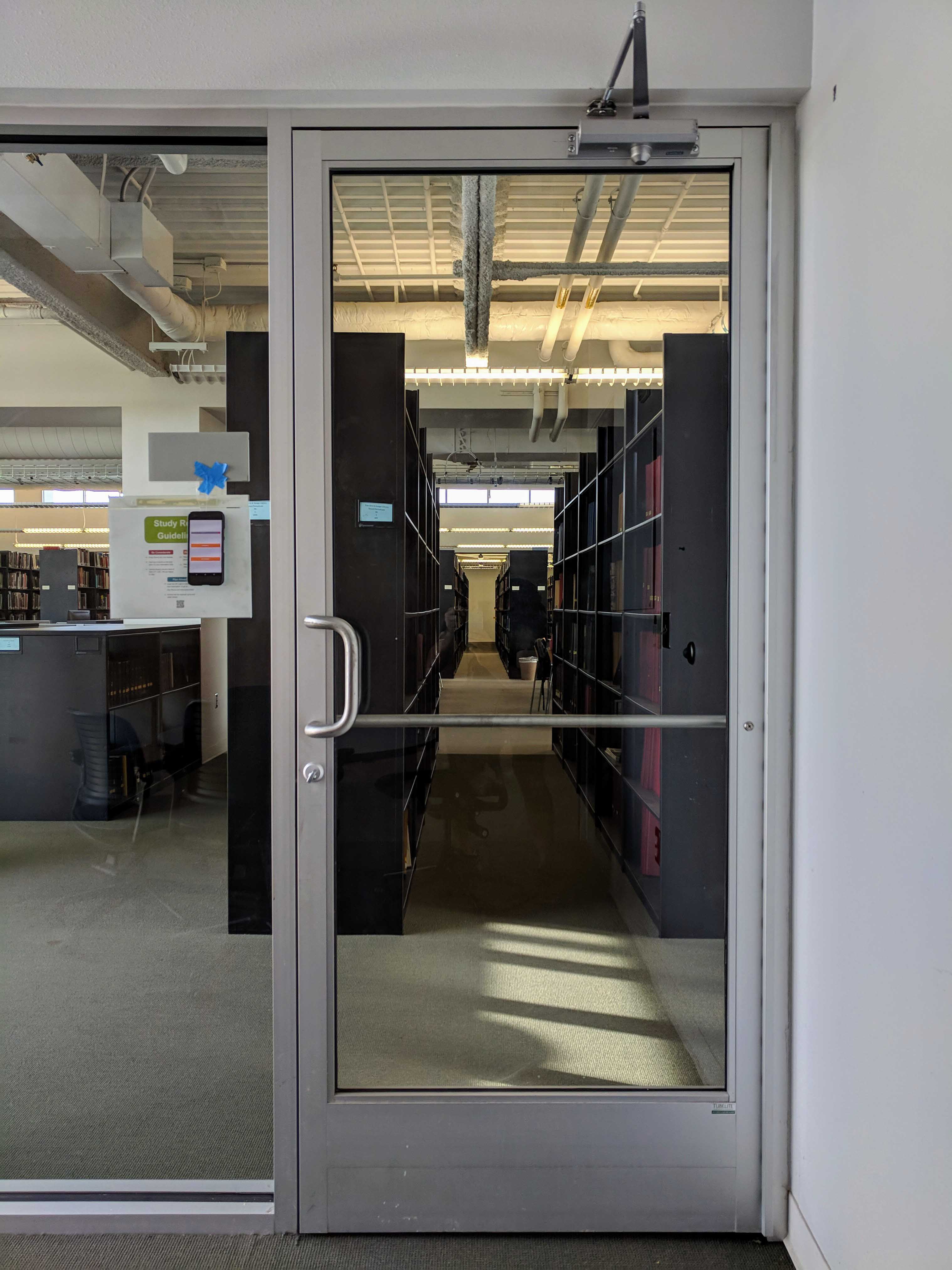}} & \multicolumn{3}{c}{\includegraphics[width=1.5in]{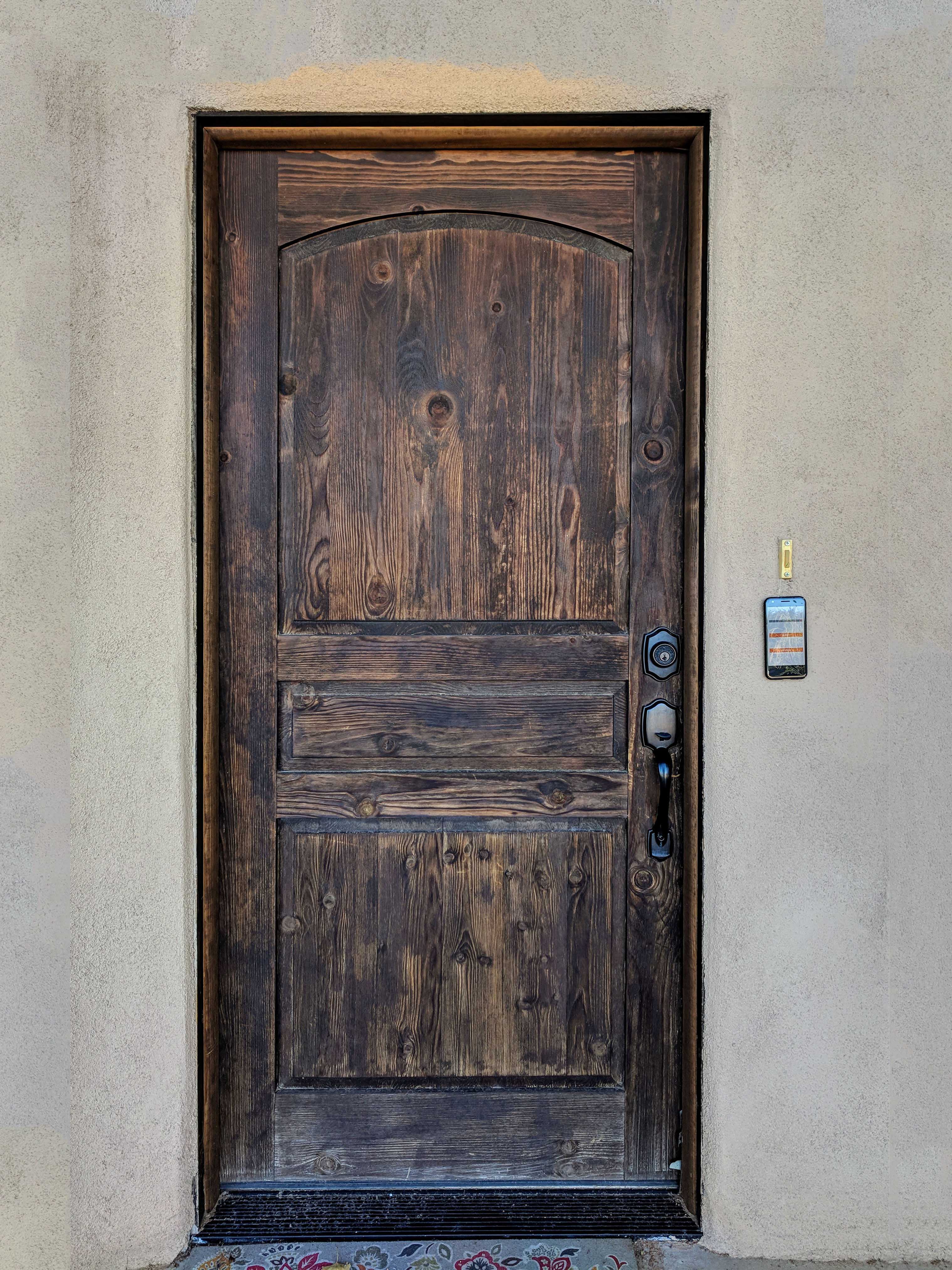}} \\
\multicolumn{2}{c}{\includegraphics[width=.95in]{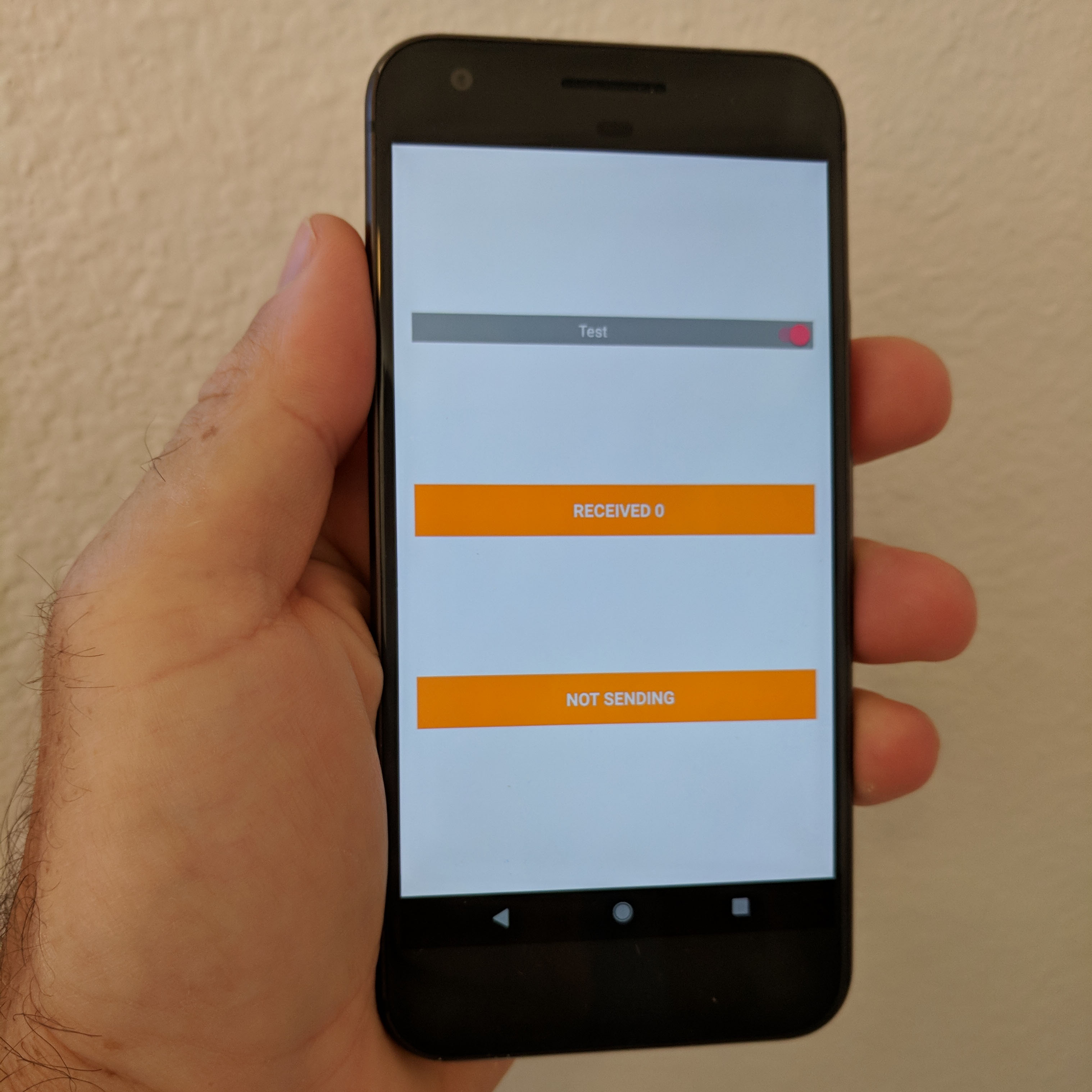}} & \multicolumn{2}{c}{\includegraphics[width=.95in]{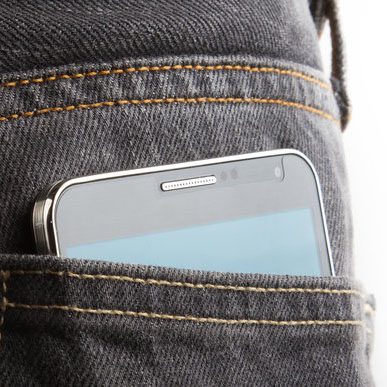}} & 
\multicolumn{2}{c}{\includegraphics[width=.95in]{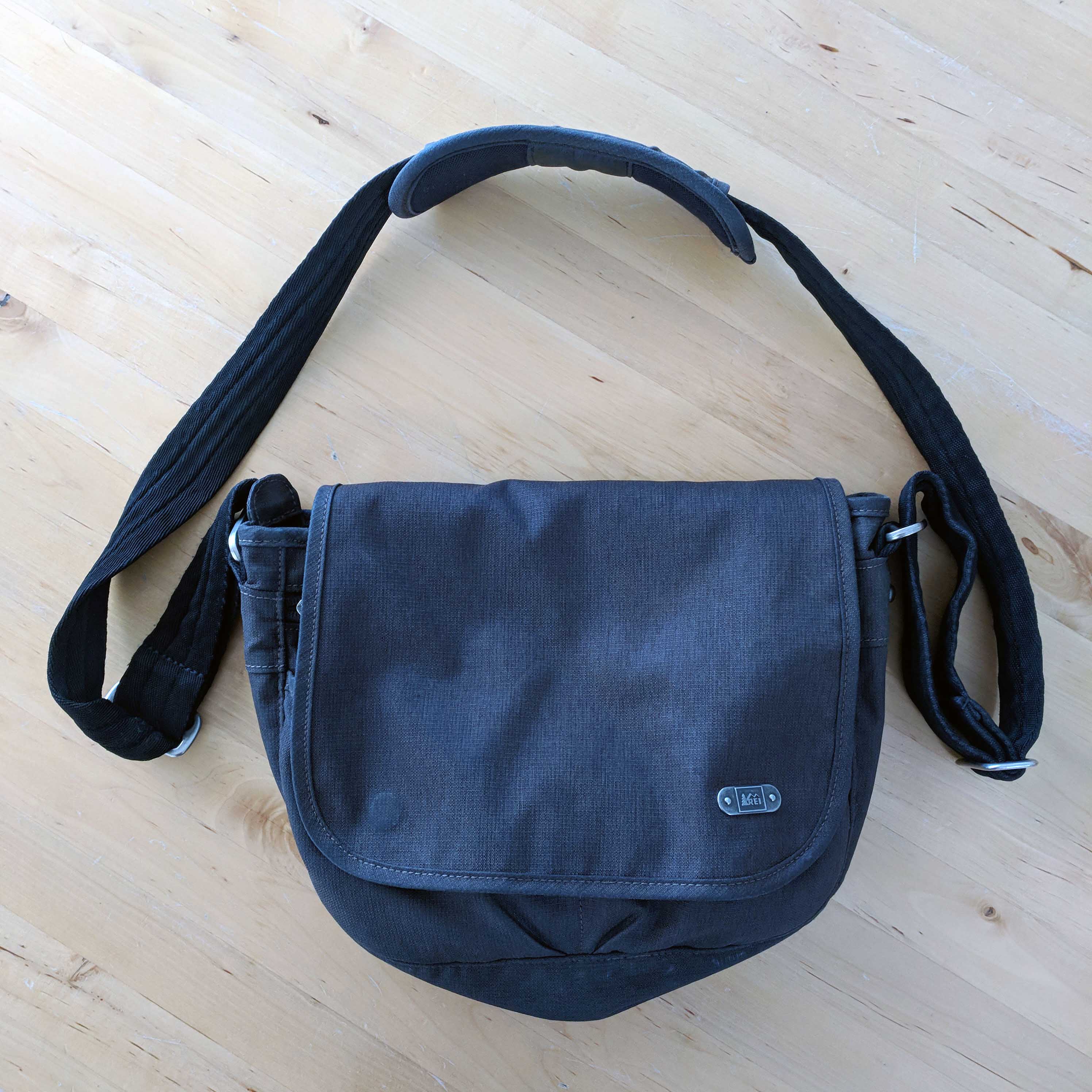}}
\end{tabular}
\caption{We evaluated the system in different environments (top) and contexts, with the mobile device held in hand, in a pocket, or placed in a bag (bottom).}
\label{fig:eval}
\end{figure}

We evaluated effectiveness of our approach for detecting a user's position relative to a threshold, examining a range of conditions, including different device contexts and threshold types. We tested the following conditions:

\textbf{Environment} (2): We compared performance for a \emph{home} entryway and an internal \emph{office} door (Figure \ref{fig:eval}, top).

\textbf{Door state} (2): While intended to work when a space is enclosed, we tested with doors both \emph{shut} and \emph{open}.

\textbf{Fixed device location} (2): A fixed device (Figure \ref{fig:eval}, top) is either \emph{internal} or \emph{external} (e.g., in the office vs. hallway).

\textbf{User location} (2): The user was in the \emph{same space} as the fixed device or in the \emph{other space}.

\textbf{Distance} (3): The user was \emph{10 ft}, \emph{2 ft}, or \emph{0 ft} from the threshold. We tested the  \emph{0 ft} distance only with the door shut and on the other side from the fixed device.

\textbf{Mobile Context} (3): The mobile was held in \emph{hand}, in a \emph{pocket}, or in a standard canvas \emph{bag} (i.e., purse). (Figure \ref{fig:eval}, bottom)

\textbf{Configuration} (2): The mobile acted as a beacon (\emph{mobile beacon}) or a receiver (\emph{mobile receiver}).

\subsection{Setup}
We performed tests using two Google Pixel (v1) devices, one fixed and mounted next to the door near its handle (see Figure \ref{fig:eval}) and one mobile. We set the volume level of both devices as high as possible such that no audible artifacts were detected (for the devices used, this was volume level 11/15). In each test, the beacon application sent a simple hexadecimal message (``0xaa'') 20 times, and the number of messages ``heard'' by the receiving device is reported. We collected a total of 4,320 samples (20 samples x 216 condition-combinations).


\subsection{Results}

\begin{figure}[t]
\centering
\begin{tabular}{c} 
\includegraphics[width=3.25in]{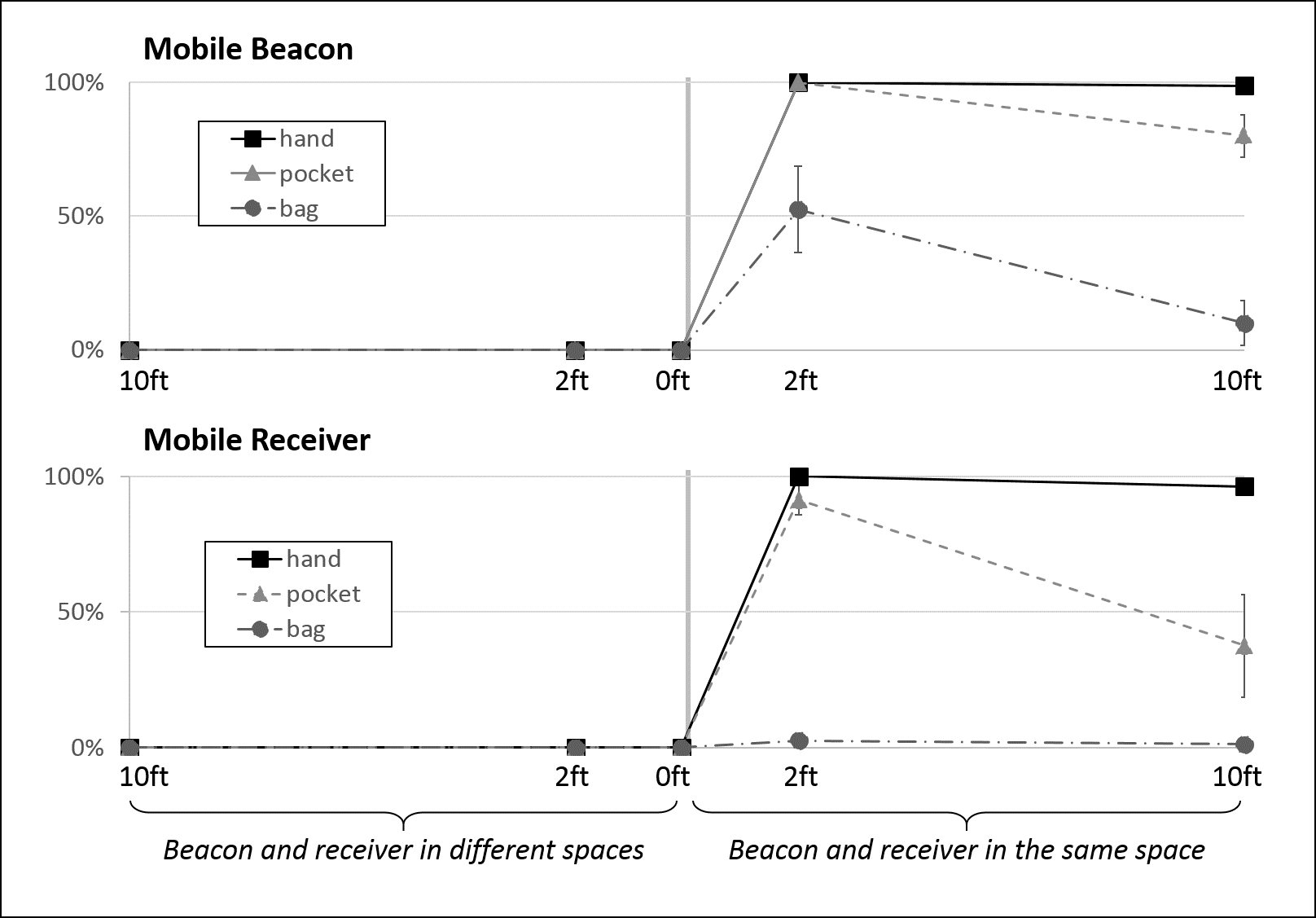}
\end{tabular}
\caption{Percent of messages received at different distances (in feet) from a closed door. in different usage contexts with the mobile device configured as a \emph{beacon} (top) and \emph{receiver} (bottom). In the ideal case, no messages are received from the other space (left side), and 100\% of messages are received in the same space (right side). }
\label{fig:results}
\end{figure}

\begin{figure}[t]
\centering
\begin{tabular}{c} 
\includegraphics[width=3.35in]{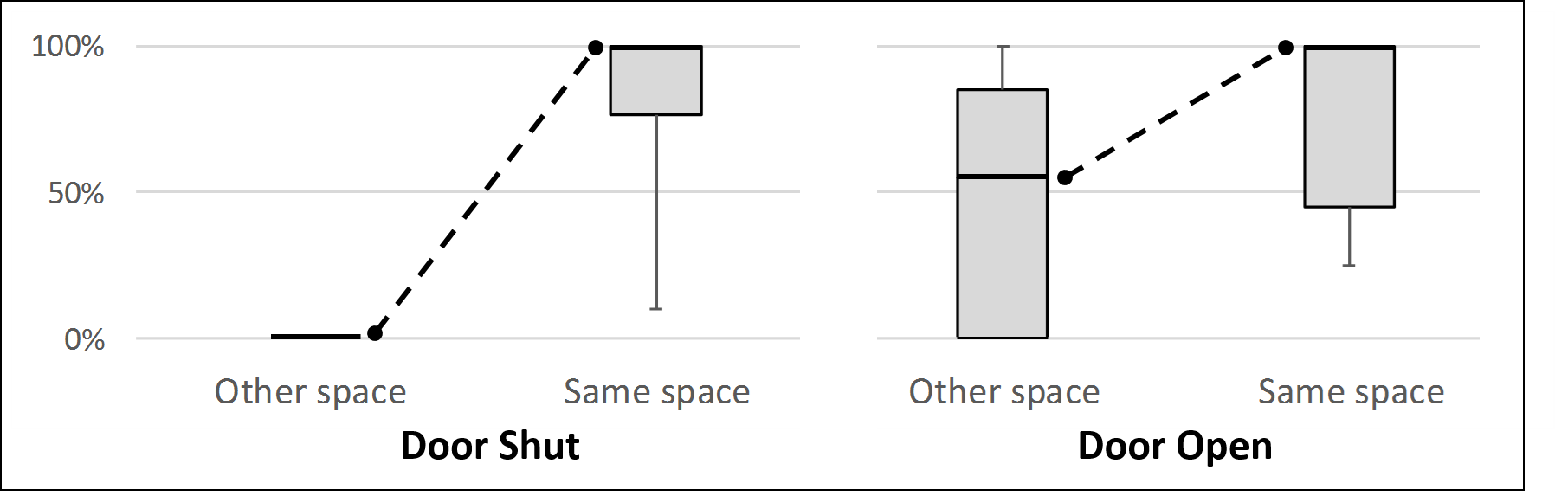}
\end{tabular}
\caption{Percent messages received for the door shut and door open conditions when the device shares the same space as the fixed device versus the other space. Illustrates how, with the door open, messages leak from the other space.}
\label{fig:results_door}
\end{figure}

Figure \ref{fig:results} shows the percent of messages received at different distances from a closed door, in different usage contexts (hand, pocket, and bag), when the phone is configured as a beacon (top) or a receiver (bottom). 
The results show that our approach works well, particularly when the phone is in the user's hand or pocket. 87\% of messages (1110 of 1280) were correctly received from a beacon in the \emph{same space} and, importantly, when the door is shut, not a single message was incorrectly received from a beacon in the other space (0/1440). In other words, when the door is shut, our system produces no false positives. We should note, however, that when the phone is inside a bag, performance is only adequate when standing near the door and the phone is configured as a beacon. 
There was no significant difference in performance between the office and home environments.

As can be seen in Figure \ref{fig:results_door} left, our system is able to distinguish extremely well standing next to, but on opposite sides of a closed door---our key use-case. However, not surprisingly, when standing near an open door (Figure \ref{fig:results_door} right), the system picks up inaudible messages from the other space (in our tests, 171 of 640 messages, or 27\%), particularly when the phone is held in the user's hand and is configured as a beacon.

Finally, we observed that more messages were received when the mobile device acted as a beacon (\emph{mobile beacon}) than as a receiver. 
As mentioned above, when placed in a bag, the mobile device acting as a beacon results in a higher percent of messages picked-up \emph{correctly} near the door (46\%) than when the phone acts as a receiver (only 3\%). However, when the door is open, a higher percent of messages ``leak'' \emph{incorrectly} in mobile beacon configuration than in a mobile receiver configuration (27\% vs. 8\%). We discuss this trade-off later. 
\section{Usage Scenarios}

We envision the use of our solution in a variety of different applications in two primary categories: controlling access to services and controlling access to spaces.

\subsection{Controlling access to services}
\emph{Secure document access.}\quad
Previous work has used precise indoor localization to ensure that users have access to certain sensitive documents (e.g., financial, medical) only while they are at a particular private location. This is often accomplished using GPS and Geofencing or similar techniques \cite{Doclock}. However, these techniques provide only coarsely-defined secure areas, allowing malignant parties to potentially access documents on the boundaries. Similarly, LocAssure \cite{biehl_youre_2015} proposed to provide assurances of a user's reported location, but their underlying system still depends on the accuracy of the (BLE) localization infrastructure. By comparison, our RF- and US-based approach can allow users to define much more fine-grained secure areas with much lower risk of adversarial document access.

\emph{Improved location-based notification.}\quad
Location-based reminders and notifications have been demonstrated in both outdoors \cite{sohn_placeits} and indoors \cite{lin_location-based_2012} settings. However, with the limited accuracy of RF-based localization, applications run the risk of delivering notifications at incorrect moments or miss otherwise useful opportunities. For example, due to the low accuracy of Bluetooth-based localization, Cambo \textit{et al.} \cite{cambo_breaksense:_2017} relied on changes in physical activity recognized by a smartwatch to trigger notification, which resulted in notification lag. Our solution would help ensure that location-based notifications arrive appropriately.

\subsection{Controlling access to space}
\emph{Smarter smart locks.}\quad
Current smart locks \cite{august,fridaylabs,ottoCNET,danalock} make use of Geofencing to automatically unlock the door when a user's device is in proximity of the lock. However, such locks will typically rely on several assumptions that can lead to errors.
For example, after automatically unlocking the door for the user, the system may incorrectly assume that the opening and then closing of the door means that the user is now inside the home when, in fact, they stayed outside, leading to an incorrect system state.
By adding our solution to the internal-facing plate of the smart lock, for example, the system will correctly sense whether the user has entered their home. Based on the results presented above, to support a phone carried in a bag or purse, the lock should be configured as a receiver.

\begin{figure}[tb]
\centering
\begin{tabular}{cc}
\includegraphics[width=1.5in]{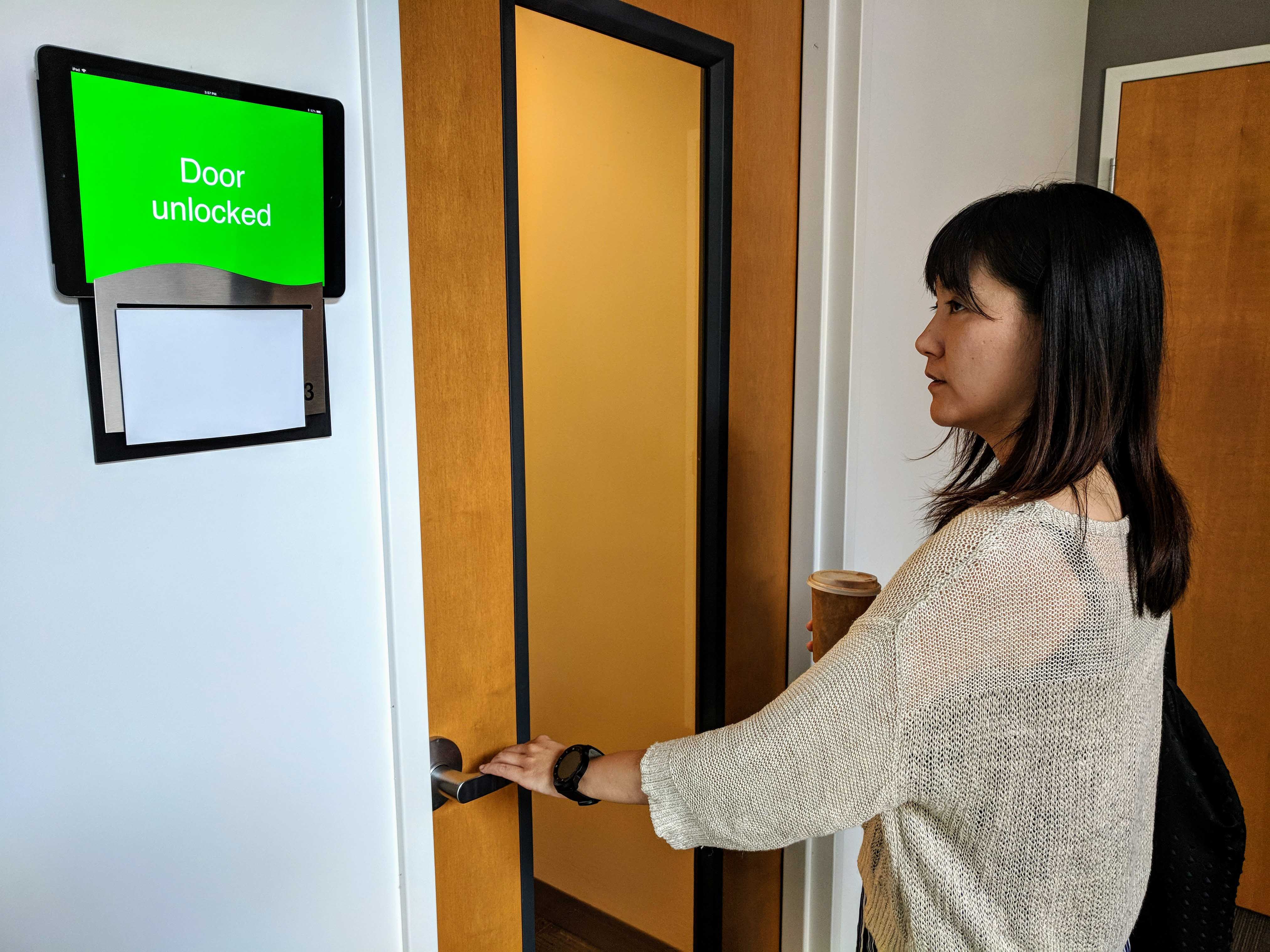} & \includegraphics[width=1.5in]{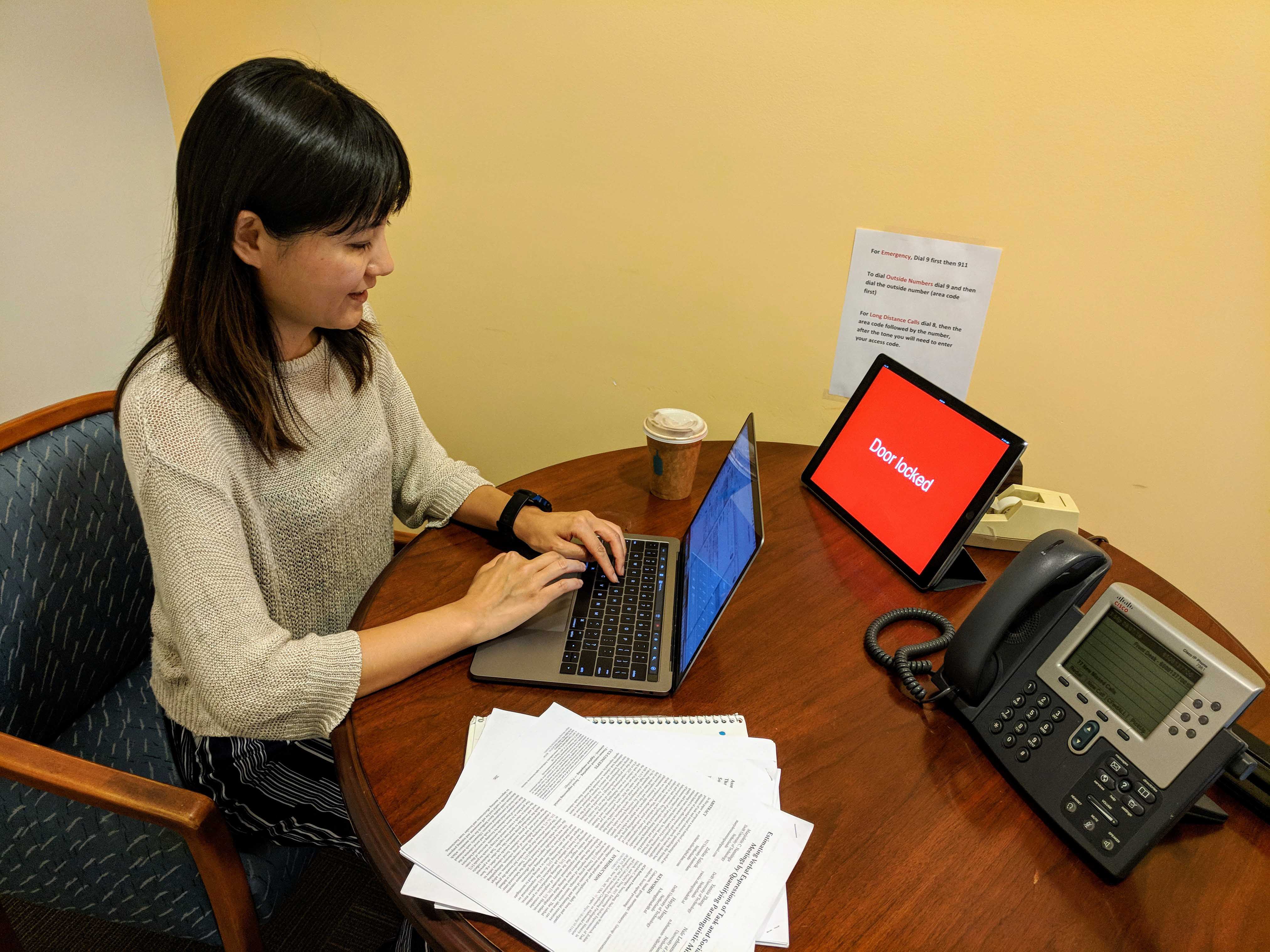} \\
\end{tabular}
\caption{Use-case illustration: Controlling access to private, rentable workspaces. A scheduling and access-control system uses inaudible audio to determine whether to grant access to a user at a particular time.}
\label{fig:human}
\end{figure}

\emph{Private, rentable workspaces.}\quad
``Workbooths'' such as Jabbrrbox \cite{jabbrrbox}, Zenbooth \cite{zenbooth}, and others, placed in public areas, allow users to schedule and pay to have a temporary, private workspace. Such rentable workspaces, which combine scheduling and access control, would benefit from correct knowledge of whether a specific user is inside or outside their scheduled space. 
The scheduling and access-control systems would be able to, for example, determine whether a user's scheduled time is active, granting them repeated access while preventing access from others (see Figure \ref{fig:human}). If multiple workspaces are located in close proximity (preventing RF-based localization from pinpointing a user's location), our solution's ability to broadcast encrypted, identifying messages through multiple ultrasonic channels would ensure that users have access only to the spaces they have scheduled.

\section{Discussion}
The evaluation demonstrated the viability of using ultrasonic audio as a low-cost solution to determine whether a user is on one side of a threshold or another---important for location- and proximity-aware smart applications. 
Performance when the phone was in a bag was lower. For this reason, in future work we plan to implement and evaluate our solution using a smartwatch, since it is less likely to be obstructed.

The evaluation also highlighted an important trade-off when deploying this solution. Specifically, in order to support detection when the user carries their phone in a bag (i.e., avoid false negatives), our tests suggest that the phone needs to be configured as a beacon and fixed devices as receivers. However, the tests also show that when the door is open and a phone, configured as a beacon, is held in hand, messages are more likely to be incorrectly received from the other space (false positives). While in some cases this trade-off could be resolved through simple door-state sensing (e.g., magnetic sensors), we demonstrate that the configuration of beacons and receivers should be considered based on the application to optimize for fewer false positives or false negatives. 

Finally, a critical area for location-aware applications in general, and for applications that rely on acoustic sensing in particular, is the threat to users' privacy (\textit{cf.}, \cite{chitkara_does_2017,klasnja_exploring_2009}). Indeed, in addition to performance trade-offs, the two configurations of our solution may have different implications for privacy. In the first configuration,  the mobile phone's microphone is opened once in RF range of a beacon, listening for ultrasonic phrases. This requires the user to grant the application access to the phone's microphone---something users may be hesitant to do. Still, a key benefit of using the phone as a receiver is that sensing can be performed locally on the phone without disclosing information about the user.
The second configuration, on the other hand, requires stationary microphones to be present and listening in the environment. Through the promise of only listening to ``wake words'', stationary, always-listening devices such as Amazon Alexa Echo \cite{noauthor_amazon_2015} and Google Home \cite{noauthor_google_2016} have become commonplace in recent years, suggesting that users may be comfortable with this configuration (although research has shown various possible risks associated with these devices \cite{arp_privacy_2017}). Always-listening devices are, however, only beginning to find acceptance in workplace environments \cite{alexa_for_business}. We argue that considerations of privacy and performance must both be taken into account when implementing and deploying solutions such as the one described in this paper.

\section{Conclusions}
We presented a simple and cost-effective solution for helping RF-based localization and proximity systems overcome the challenge of determining positions near thresholds such as walls and doors. Our approach takes advantage of the limited propagation of sound waves between enclosed spaces. Our evaluation showed high performance of the solution, especially when the user's phone is held or in a pocket. Finally, we discussed performance and privacy trade-offs that must be considered based on the intended target application.


%
%
%
%
%

\balance{}

\bibliographystyle{SIGCHI-Reference-Format}
\bibliography{main}

\end{document}